\documentclass[prl,twocolumn,superscriptaddress]{revtex4}

\usepackage{graphicx}
\usepackage{graphics}      
\usepackage{bm}    
\usepackage{amsmath}     
\usepackage{amsfonts}
\usepackage{amssymb}
\usepackage{latexsym}  

\tighten

\begin{document}

\title{Stronger no-cloning, no-signalling and conservation of 
quantum information }
\newcounter{count}
\author{I. Chakrabarty}
\email{indranilc@indiainfo.com} \affiliation{Heritage Institute of
Technology, Kolkata, India} \affiliation{Bengal Engineering and
Science University, Howrah-711103, West Bengal, India}
\author{A. K. Pati}
\email{akpati@iopb.res.in}
\affiliation{Institute of Physics, Bhubaneswar-751005, Orissa,India}
\author{S. Adhikari}
\email{satyyabrata@yahoo.com}\affiliation{Bengal Engineering and
Science University, Howrah-711103, West Bengal, India}
\date{\today}
\begin{abstract}
It is known that the stronger no-cloning theorem and the
no-deleting theorem taken together provide the permanence property of
quantum information. Also, it is known that the violation of the 
no-deletion theorem would imply signalling. Here, we show that the violation 
of the stronger no-cloning theorem could lead to signalling. Furthermore, 
we prove the stronger no-cloning theorem from the conservation of
quantum information. These observations imply that the permanence property of
quantum information is connected to the no-signalling and the conservation of 
quantum information.
\end{abstract}
\maketitle
{\bf Introduction:}
In recent years it is of fundamental importance to know 
various differences between classical and quantum information.
Many operations which are feasible in digitized information
become impossibilities in quantum world. Unlike classical
information, in quantum information theory, we cannot
clone and delete an arbitrary quantum states which are now 
known as the no-cloning and the no-deleting theorems \cite{wz,yuen,pb}. In
addition to these two there are many other theorems
which come under this broad heading of `no-go theorems' 
\cite{bhw,gp,pb1,ps,zzy}.
In particular, we have the `no-flipping' \cite{bhw,gp}, 
the `no-self replication' \cite{pb1}, the `no-partial erasure' \cite{ps}, the 
`no-splitting' \cite{zzy}, and few other theorems in 
the literature.  
Moreover, the no-cloning, the no-conjugation,
and the no-complementing theorems have been unified under a name 
`General Impossible
operations' in quantum Information theory \cite{pati}. 

Now, we know that these
impossible theorems can also be derived from the fundamental
principles like impossibility of signalling and non-increase of
entanglement under local operations for closed systems 
\cite{gisin,hardy,pati1,pati2,chat}. For example, 
the no-deleting theorem was  
established from the principle of no-signalling between two spatially separated
parties \cite{pati2}. The no-flipping theorem has been obtained from 
the no-signalling and the non-increase of entanglement under LOCC \cite{chat}. 
Recently, it was shown that the no-cloning
and the no-deletion theorems follow from the conservation of
quantum information \cite{horo}. In strengthening the no-cloning theorem it was
proved that in order to make a copy of a set of non-orthogonal
states one must supply full information. This is called as the
stronger no-cloning theorem \cite{jozsa}. The stronger no-cloning and the
no-deleting theorems taken together implies a permanence of quantum
information. Since the deletion operation implies signalling one may
wonder whether violation of the stronger no-cloning theorem in 
quantum information could also lead to signalling. If yes, then that 
would mean that violation of permanence property of quantum information could 
lead  to signaling. In this work we show 
that indeed principles like no-signalling and
conservation of entanglement under LOCC establish the permanence of 
quantum information. 

Before going into the details, first of all, we should keep
in mind that we are treating information not only as knowledge but
also as a physical quantity \cite{land}. In fact, there are two opposing
concepts regarding information: (i) subjective, according to this
information is nothing but knowledge obtained about a system after
interacting with it, and (ii) objective, which tells us that 
information is a physical quantity. Now we must understand what
actually we mean by the term permanence of quantum information.
It refers to the fact that neither we could create nor destroy 
quantum information.
From the objective viewpoint quantum information resemble like
energy which already existed before any physical process has taken
place and will continue to exist even after any physical process.
Permanence of quantum information is something which is akin to this. 
It is entirely quantum mechanical in nature and there is no classical 
analogue.

Recall that the no-cloning theorem states that 
if  $\{|\psi_i\rangle\}$ is a set of
pure non-orthogonal states, then there is no unitary 
operation that can achieve the transformation
$|\psi_i\rangle \rightarrow |\psi_i\rangle|\psi_i\rangle$ \cite{yuen}.
However, the stronger no-cloning theorem states that 
if  $\{|\psi_i\rangle\}$ is a set of
pure states containing  no orthogonal pair, then 
the transformation 
\begin{eqnarray}
|\psi_i\rangle |a_i \rangle \rightarrow |\psi_i\rangle|\psi_i\rangle
\end{eqnarray}
is possible iff $|a_i \rangle \rightarrow |\psi_i\rangle$.  
That is the  second copy $|\psi_i\rangle$ can
always be generated from the supplementary information alone, 
independent of the first copy. Thus in effect cloning of
$|\psi_i\rangle$ is possible only when the second copy is provided
as an additional
input. This is known as the `stronger no-cloning theorem' \cite{jozsa}.

Let us briefly recapitulate the stronger no-cloning theorem. 
Consider a physical process described by the equation
\begin{eqnarray}
|\psi_i\rangle|0\rangle|\alpha_i\rangle|C\rangle\rightarrow 
|\psi_i\rangle|\psi_i\rangle|C_i\rangle, 
\end{eqnarray}
where $|\psi_i\rangle$ is the copy which is to be cloned and
$|0\rangle$ is the blank state. Here $|C\rangle$ is the
initial state of the environmental space and $|\alpha_i\rangle$ is
the state which already existed in some other part of the world.
Now $|C_i\rangle$ is the output state of the two register which
initially contained $|\alpha_i\rangle, |C\rangle$. Since we have
included the environmental space in this physical process,  we can
consider the transformation (2) as a unitary transformation.
Hence from unitarity, the sets $\{ |\alpha_i\rangle \}$ and $\{
|\psi_i\rangle, |C_i\rangle \}$ have equal matrices of inner
product. Then, $\{ |\alpha_i\rangle \}$ and $\{
|\psi_i\rangle, |C_i\rangle]\}$ are unitarily equivalent and in
advance we can say that $|\psi_i\rangle$ is generated from
$|\alpha_i\rangle$ (discarding the ancillary system
$|C_i\rangle$). This is  known as the `Stronger no-cloning' theorem 
\cite{jozsa}.

Now if we look into the process of deletion of non-orthogonal states 
we find that the permanence of quantum information plays an important 
role. Suppose
we are provided with two copies $|\psi_i\rangle |\psi_i\rangle$ of a
quantum state and we wish to delete one copy by some physical operation 
(a completely positive, trace preserving map) 
like
\begin{eqnarray}
|\psi_i\rangle|\psi_i\rangle\rightarrow|\psi_i\rangle|0\rangle.
\end{eqnarray}
Such a physical operation will corresponds to a linear operation
if we include ancilla and would be given by 
\begin{eqnarray}
|\psi_i\rangle|\psi_i\rangle|A\rangle\rightarrow 
|\psi_i\rangle|0\rangle|A_i\rangle.
\end{eqnarray}
The no-deleting principle tells us that the second copy
$|\psi_i\rangle$ can never be deleted in the sense that
$|\psi_i\rangle$ can always be obtained from $|A_i\rangle$.

Thus, we see in both deletion and cloning we obtain a similarity:
the first copy provided no assistance in construction of the
second copy and in deletion the first copy has not provided any
assistance in deleting the second copy. Thus, we can say that the
quantum information has some property called permanence which
tells us that the creation of copies will only be possible if we
import information from some other part which already existed and
deletion can only be possible by exporting information to some
other part where it will continue to exist.

The aim of this work is basically two fold. In the first
part we will show that negating the stronger no-cloning theorem
will lead to signalling. Since we already know that negating the
no-deletion theorem will
lead to signalling \cite{pati2}. Therefore, together with our proof we can
say that the violation of the permanence property of
quantum information will lead to signalling. Alternately, 
we can say that the no-signalling leads to permanence of
quantum information. In the second part, we will show that the
violation of the stronger no-cloning theorem will lead to the
violation of the principle of conservation of entanglement under
LOCC (conservation of information). This clearly indicates that the
principle of conservation of entanglement under LOCC implies the
permanence of quantum information.

{\bf Stronger no-cloning from no-signalling:}
Suppose we have two singlet states $ |\chi\rangle $, $ |\xi\rangle
$ shared by two distant parties Alice and Bob. Since the singlet
states are invariant under local unitary operations, it remains
same in all basis. The states are given by,
\begin{eqnarray}
|\chi\rangle=\frac{1}{\sqrt{2}}(|\psi_1\rangle|\overline 
{\psi_1}\rangle-|\overline{\psi_1}\rangle|\psi_1\rangle)\nonumber\\
=\frac{1}{\sqrt{2}}(|\psi_2\rangle|\overline 
{\psi_2}\rangle-|\overline{\psi_2}\rangle|\psi_2\rangle)\\
|\xi\rangle=\frac{1}{\sqrt{2}}(|\alpha_1\rangle|\overline 
{\alpha_1}\rangle-|\overline{\alpha_1}\rangle|\alpha_1\rangle)\nonumber\\
=\frac{1}{\sqrt{2}}(|\alpha_2\rangle|\overline 
{\alpha_2}\rangle-|\overline{\alpha_2}\rangle|\alpha_2\rangle), 
\end{eqnarray}
where $\{|\psi_1\rangle, |\overline{\psi_1}\rangle \}$, $\{
|\psi_2\rangle, |\overline{\psi_2}\rangle \}$ are two sets of
mutually orthogonal spin states or polarizations in case of
photons (qubit basis) for the first entangled pair given by (5)
shared between Alice and Bob. Similarly $\{|\alpha_1\rangle,
|\overline{\alpha_1}\rangle \}$, $\{ |\alpha_2\rangle,
|\overline{\alpha_2}\rangle \}$ are two sets of mutually
orthogonal spin states (qubit basis) for the second entangled pair
(6). Let us write the combined state of the system in an arbitrary 
qubit basis.
\begin{eqnarray}
|\chi\rangle|\xi\rangle=\frac{1}{2}(|\psi_i\rangle|\overline 
{\psi_i}\rangle-|\overline{\psi_i}\rangle|\psi_i\rangle)(|\alpha_i\rangle 
|\overline{\alpha_i}\rangle-|\overline{\alpha_i}\rangle|\alpha_i\rangle)
\end{eqnarray}
where $i=1,2$. Note that whatever measurement
Alice does, Bob does not learn the results and his description will
remain as that of a completely random mixture , i.e., 
$\rho_{B}=\frac{I}{2}\otimes \frac{I}{2}$. In other words we can
say that the local operations performed on her subspace has no
effect on Bob's description of his states.

Now we look out for the equivalent mathematical expression for the
statement like `negating the stronger no-cloning theorem'. Let us define 
\begin{eqnarray}
&&|\psi_i\rangle|\alpha_i\rangle|C\rangle\rightarrow 
|\psi_i\rangle|\psi_i\rangle|C_{i1}\rangle{}\nonumber\\&&
|\overline{\psi_i}\rangle|\overline{\alpha_i}\rangle|C\rangle\rightarrow 
|\overline{\psi_i}\rangle|\overline{\psi_i}\rangle|C_{i2}\rangle{}\nonumber\\&&
|\psi_i\rangle|\overline{\alpha_i}\rangle|C\rangle \rightarrow 
|\phi_i\rangle{}\nonumber\\&&
|\overline{\psi_i}\rangle|\alpha_i\rangle|C\rangle\rightarrow 
|\overline{\phi_i}\rangle, 
\end{eqnarray}
where $i=1,2$ and $|\phi_i \rangle$, $|\overline{\phi_i} \rangle$ are 
some general states of the combined system.
The above equations corresponds to a situation where
we are negating stronger no-cloning theorem by allowing the
cloning of the input state $ |\psi_i\rangle $ to be dependent on the
register $ |\alpha_i\rangle $. Now, suppose Bob does some local
operation on his system by attaching ancilla  $|C\rangle$ and
applying the transformations given by (8). As a result, 
the combined system given by (7) takes the form
\begin{eqnarray}
&&|\chi\rangle|\xi\rangle|C\rangle\rightarrow{}\nonumber\\&&
\frac{1}{2}[|\psi_i\rangle |\alpha_i\rangle (
|\overline{\psi_i}\rangle |\overline{\psi_i}\rangle|C_{i2}\rangle
)+ |\overline{\psi_i}\rangle|\overline{\alpha_i}\rangle
(|\psi_i\rangle |\psi_i\rangle|C_{i1}\rangle){}\nonumber\\&&
+|\psi_i\rangle
|\overline{\alpha_i}\rangle(|\overline{\phi_i}\rangle)+ 
|\overline{\psi_i}\rangle|\alpha_i\rangle(|\phi_i\rangle)].
\end{eqnarray}
Now if Alice finds her particles in the basis $\{|\psi_1\rangle,
|\overline{\psi_1}\rangle \}$ and $\{|\alpha_1\rangle,
|\overline{\alpha_1}\rangle \}$ , the reduced density matrix
describing the Bob's system is given by
\begin{eqnarray}
&&\rho_B=\frac{1}{4}[|\psi_1\psi_1\rangle\langle\psi_1\psi_1|+ 
|\overline{\psi_1}\overline{\psi_1}\rangle\langle\overline 
{\psi_1}\overline{\psi_1}|
{}\nonumber\\&&+|\phi_1\rangle\langle\phi_1|+ 
|\overline{\phi_1}\rangle\langle\overline{\phi_1}|].
\end{eqnarray}
Now if Alice finds her particles in the basis $\{|\psi_2\rangle,
|\overline{\psi_2}\rangle \}$ and $\{|\alpha_2\rangle,
|\overline{\alpha_2}\rangle \}$, the reduced density matrix of
Bob's system is given by
\begin{eqnarray}
&&\rho_B=\frac{1}{4}[|\psi_2\psi_2\rangle\langle\psi_2\psi_2|+ 
|\overline{\psi_2}\overline{\psi_2}\rangle\langle\overline{\psi_2} 
\overline{\psi_2}|
{}\nonumber\\&&+|\phi_2\rangle\langle\phi_2|+ 
|\overline{\phi_2}\rangle\langle\overline{\phi_2}|].
\end{eqnarray}
Since the statistical mixture in (10) and (11) are different, this
would allow Bob to distinguish in which basis Alice has
performed measurement, thus allowing for super luminal
signalling.
Thus, we see that the violation of stronger no-cloning theorem will
lead to signalling. It had already been seen that the violation of
the no-deleting theorem will lead to signalling \cite{pati2}, henceforth we
can conclude that the violation of permanence property of quantum
information will make super luminal signalling possible.

{\bf Stronger no-cloning theorem from conservation of information:}
Now we will show the principle of conservation of entanglement (
conservation of information in a closed system) under LOCC will
imply the stronger no-cloning theorem in quantum information. 
This will prove that if 
we violate the permanence of quantum information we will also violate the
principle of conservation of entanglement (information). 

Now negating the stronger no-cloning theorem is equivalent of
rejecting the unitary equivalence of two sets of pure states
$\{|\alpha_i\rangle\}$ and $\{ |\beta_i\rangle \}$ (say) having equal
matrices of inner product i.e $\langle
\alpha_i|\alpha_j\rangle=\langle\beta_i|\beta_j\rangle$ \cite{jozsa}. In
other words we can say that there exists an unitary operator
operator $U$ which describe the physical process (2) but violate
the condition $\langle
\alpha_i|\alpha_j\rangle=\langle\beta_i|\beta_j\rangle$.
Now we look out for the equivalent mathematical expression for the
statement like `negating the stronger no-cloning theorem'. Consider the 
following transformations 
\begin{eqnarray}
|\psi_i\rangle|0\rangle|\alpha_i\rangle|C\rangle\rightarrow 
|\psi_i\rangle|\psi_i\rangle|C_i\rangle\\
|\psi_j\rangle|0\rangle|\alpha_j\rangle|C\rangle\rightarrow
|\psi_j\rangle|\psi_j\rangle|C_j\rangle
\end{eqnarray}
and let us negate the stronger no-cloning theorem, i.e., we assume that
there exists a unitary operator, for which the inner products are not
equal
\begin{eqnarray}
\langle
\alpha_i|\alpha_j\rangle\neq\langle\psi_i|\psi_j\rangle\langle
C_i|C_j\rangle.
\end{eqnarray}
Let us consider an entangled state shared by two distant parties
Alice and Bob
\begin{eqnarray}
|\Psi\rangle=\frac{1}{\sqrt{2}}[|0\rangle_A|\psi_i\rangle_B|\alpha_i\rangle_B +
|1\rangle_A|\psi_j\rangle_B|\alpha_j\rangle_B],
\end{eqnarray}
where Alice is in possession of the qubit `A' and Bob is in
possession of the qubit `B'. Now imagine that Bob is in possession of
a hypothetical machine whose action on his qubit is given by the
transformation defined in (12) and (13). He attaches two ancillas in the
state $|0\rangle_B |C\rangle_B$ and carries out the transformation 
(12) and (13).

The reduced density matrix on Alice side before the application of
hypothetical machine is given by
\begin{eqnarray}
\rho^A=\frac{1}{2}[|0\rangle\langle0|+|1\rangle\langle0| 
\langle\psi_i|\psi_j\rangle\langle\alpha_i|\alpha_j\rangle + 
\nonumber\\
|0\rangle\langle1| \langle\psi_j|\psi_i\rangle\langle\alpha_j|\alpha_i\rangle
+|1\rangle\langle1|].
\end{eqnarray}
Now after the application of the transformation defined in (12)
and (13) by Bob on his qubits the entangled state takes the form,
\begin{eqnarray}
|\Psi\rangle_{SNC}=\frac{1}{\sqrt{2}}[|0\rangle_A|\psi_i\rangle_B 
|\psi_i\rangle_B|C_i\rangle_B+ \nonumber\\
|1\rangle_A|\psi_j\rangle_B|\psi_j\rangle_B|C_j\rangle_B].
\end{eqnarray}
The reduced density matrix on Alice side by tracing out Bob's part
is given by
\begin{eqnarray}
\rho^A_{SNC}=\frac{1}{2}[|0\rangle\langle0|+|1\rangle\langle0| 
\langle\psi_i|\psi_j\rangle^2 \langle C_i|C_j\rangle \nonumber\\
+|0\rangle\langle1| \langle\psi_j|\psi_i\rangle^2 
\langle C_j |C_i\rangle + |1\rangle\langle1|].
\end{eqnarray}
Now from equation (14) we can say that equations (16) and (18) are
not identical. Since transformation defined in (12) and (13) are
local operation and there is no classical communication between
two distant parties Alice and Bob, to maintain the principle of no
signalling the density matrix on Alice's side must remain
unchanged. A different density matrix on Alice's side will imply
that the signalling has taken place.

Let us consider the entangled state given by equation (15), which
is shared by two distant parties Alice and Bob. The largest eigen
value of the reduced density matrix (16) on Alice's side is given
by
\begin{eqnarray}
\lambda^A=\frac{1}{2}+\frac{|a||b|}{2}
\end{eqnarray}
and the largest eigen value of the density matrix (18) on Alice's
side after the application transformation defined in (12) and (13)
is given by
\begin{eqnarray}
\lambda^A_{SNC}=\frac{1}{2}+\frac{|a|^2|c|}{2},
\end{eqnarray}
where $a=\langle\psi_i|\psi_j\rangle$, $b=\langle
\alpha_i|\alpha_j\rangle$ and $c=\langle C_i|C_j\rangle$.\\
Now from equation (14) it is clear that equations (19) and (20)
are not identical. This violates the principle of conservation of
entanglement under LOCC. Thus we again observe that by violating
the permanence property of quantum information we are violating
the principle of conservation of entanglement under LOCC or in
other words we can say that the principle of conservation
of entanglement (conservation of information) under LOCC implies
the permanence property of
quantum information.

{\bf Conclusion:}
In summary, we have shown that if we could negate the stronger no-cloning 
theorem then we could send signal faster than light. Thus, the stronger 
no-cloning follows from the no-signalling. Earlier, it has been shown that 
negating the no-deletion theorem also leads to signalling \cite{pati2}. 
Since the 
stronger no-cloning and the no-deleting theorems taken together 
provide permanence to quantum 
information, this in turn implies that the violation of the 
permanence property of quantum information could also lead to 
signalling condition. 
In other words the only possibility that we can import and export 
quantum information without being able to creat or destroy is consistent 
with no-signalling condition.
Also, we have proved the stronger no-cloning theorem from the 
conservation of quantum information under LOCC operation.

\vskip 1cm

{\it Acknowledgement:}
Indranil and Satyabrata acknowledge Prof B. S. Choudhuri, Head of
the Department of Mathematics, Bengal Engineering and Science
University. Indranil acknowledge Prof C. G. Chakraborti, S.N.Bose
Professor of Theoretical Physics, Department of Applied
Mathematics, University of Calcutta for being the source of
inspiration  in carrying out research. This work is done in
Institute of Physics, Bhubaneshwar, Orissa. Indranil and
Satyabrata gratefully acknowledge their hospitality
and also acknowledge Dr. P. Agarwal for having various useful
discussions. Authors also acknowledge Dr. G. P. Kar for giving
various useful suggestions
and encouragement for the improvement of the work.

\end{document}